\documentstyle[amsfonts]{article}
\input epsf

\newcommand{\bea}{\begin{eqnarray}}
\newcommand{\eea}{\end{eqnarray}}
\newcommand{\beq}{\begin{equation}}
\newcommand{\eeq}{\end{equation}}
\newcommand{\cd}{\partial}
\newcommand{\CP}{{\Bbb C}P^{1}}
\newcommand{\R}{{\Bbb R}}
\newcommand{\C}{{\Bbb C}}
\newcommand{\Z}{{\Bbb Z}}
\newcommand{\ra}{\rightarrow}

\newcommand{\ol}{\overline}
\newcommand{\vvec}{\mbox{\boldmath $v$}}
\newcommand{\sigvec}{\mbox{\boldmath $\sigma$}}
\newcommand{\tauvec}{\mbox{\boldmath $\tau$}}

\newcommand{\zv}{{\rm\bf 0}}

\title{Gravity thaws the frozen moduli of the $\CP$ lump}
\author{J. M. Speight\thanks{E-mail: {\tt mspeight@mis.mpg.de}} \\ 
Max-Planck-Institut f\"{u}r Mathematik in
den Naturwissenschaften \\ Inselstra{\ss}e 22-26, 04103 Leipzig, Germany \\
\\
I. A. B. Strachan\thanks{E-mail: {\tt i.a.strachan@maths.hull.ac.uk}} \\
Department of Mathematics\\
University of Hull\\
                Cottingham Road,
                Hull HU6 7RX,
                England}
\date{}

\begin{document}
\maketitle
\begin{abstract}
The slow motion of a self-gravitating $\CP$ lump is investigated in the 
approximation of geodesic flow on the moduli space of unit degree static
solutions $M_1$. It is found that moduli which are frozen in the absence
of gravity, parametrizing the lump's width and internal orientation, may
vary once gravitational effects are included. If gravitational coupling
is sufficiently strong, the presence of the lump shrinks physical space
to finite volume,
and the moduli determining the boundary value of the $\CP$ field thaw also.
Explicit formulae for the metric on $M_1$ are found in both the weak and strong
coupling regimes. The geodesic problem for weak coupling is studied in detail,
and it is shown that $M_1$ is geodesically incomplete. This leads to the
prediction that self-gravitating lumps are unstable.

\end{abstract}

\section{Introduction}
\label{sec:intro}
Collective coordinate approximations have proved to be an invaluable tool in
the study of topological soliton dynamics. They truncate the infinite
dimensional dynamical system of the full field theory to a finite dimensional
approximation which (hopefully) captures the important features of soliton
motion, but is more amenable to analytic effort. For field
theories of Bogomol'nyi type, the truncated system for $n$-soliton dynamics is
obtained simply by restricting the field theoretic Lagrangian to the moduli
space of degree $n$ static solutions, $M_n$ \cite{man}. A rather beautiful
feature in this case is that the reduced dynamics amounts to geodesic motion 
on $M_n$ (hence this is often called the geodesic approximation). The metric
on $M_n$ descends from the kinetic energy functional of the field theory.

The geodesic approximation has been applied in many contexts (BPS monopoles
\cite{atihit}, abelian Higgs vortices \cite{sam,iabs} and discrete sine-Gordon
kinks \cite{tdsg}, for example) with considerable success. One system for 
which the approximation encounters problems is the $\CP$ model in $(2+1)$ 
dimensions \cite{war,lee}. 
Here, a single static soliton (conventionally called a ``lump'')
has 6 parameters: 2 give its position, 3 its internal  orientation and 1 sets
its scale (the static field equation is conformally invariant). Hence $M_1$ is
6 dimensional. The metric on $M_1$ is, however, ill defined because at each 
static solution, 4 of the
6 zero modes are not normalizable, essentially due to the lump's weak
(polynomial) spatial localization. The only ones which {\em are} normalizable
generate spatial translations. It follows that 4 of the 6 moduli are frozen in
this approximation, since any change in them is impeded by infinite inertia
\cite{war}. Of these 4, the 2 which specify the boundary value of the $\CP$
valued field should be expected to be frozen. That the other 2, determining 
the lump's width and its orientation about the boundary value, are frozen is
not in good agreement with numerical simulations \cite{zak}. These suggest
that more general 
motion than simply constant velocity drift is possible. 

So the geodesic approximation for the $\CP$ model is unfortunately rather
singular. The singularity is removed if physical space $\R^2$ is replaced by
a compact Riemann surface (e.g.\ $S^2$ or $T^2$), yielding well defined and
mathematically interesting  geodesic problems \cite{sphere,torus}. However,
there is no obvious physical justification for such an assumption.

In this note we wish to describe a more natural means of removing the 
singularity: namely including gravitational self interaction. The moduli
space $M_1$ of static self-gravitating lumps is already explicitly known,
due to work of Comtet and Gibbons \cite{comgib}. 
The geometric distortion of physical space introduced by gravity
is just sufficient to regulate the divergent integrals encountered in
\cite{war} when evaluating $g$, the metric on $M_1$. One expects, therefore,
that the previously frozen moduli thaw once gravitational effects are taken
into account. By restricting the Einstein-Hilbert-$\CP$ action to fields which
are at all times in $M_1$, we shall demonstrate that this is indeed the
case. The degree of thawing depends on the strength of the coupling to
gravity: for small coupling, physical space has infinite 
volume and the boundary
value of the $\CP$ field remains frozen, while above a critical coupling, space
has only finite volume 
 and all 4 of the previously frozen moduli thaw. We shall derive 
explicit expressions for $g$ in both cases, and investigate one-lump dynamics
with weak gravitational coupling in detail.

\section{The moduli space}
\label{sec:moduli}

The Einstein-Hilbert-$\CP$ action is
\beq
I[W,\eta]=\int d^3x\, \sqrt{|\eta|}\left[
-\frac{S_\eta}{16\pi G}+2\mu^2\eta^{\alpha\beta}
\frac{\cd_\alpha W\cd_\beta\ol{W}}{(1+|W|^2)^2}\right],
\eeq
where $\eta$ is the Lorentzian metric on spacetime (assumed, for the moment,
to be diffeomorphic to $\C\times\R$), $|\eta|=\det(\eta_{\alpha\beta})$,
$S_\eta$ is the scalar curvature of $\eta$, $W$ is an inhomogeneous coordinate
on $\CP$ (or equivalently a complex stereographic coordinate on $S^2$), $G$
is Newton's constant and $\mu$ is a coupling constant. One may obtain static
degree 1 (degree here meaning the topolgical degree of the mapping 
$W:\C\cup\{\infty\}\ra
\CP$) solutions of the corresponding field equations by imposing the metric 
ansatz
\beq
\eta=dt^2-\Omega^2(z,\ol{z})dz\, d\ol{z},
\eeq
$z$ being a complex coordinate on physical space. The static field equation for
$W$ is conformally invariant, hence independent of $\Omega$. It follows that
all the ``flat space'' solutions of the non-gravitating model
(namely $W=$ rational map in $z$ or $\ol{z}$) survive. In particular, the most
general degree 1 solution is 
\beq
W(z)=\frac{a_{11}z+a_{12}}{a_{21}z+a_{22}}=:
\left(\begin{array}{cc}
a_{11} & a_{12} \\
a_{21} & a_{22} \end{array}\right)\odot z=M\odot z
\eeq
with $\det M\neq 0$. Note that the identification of a particular rational map
$W$ with a $GL(2,\C)$ matrix $M$ is nonunique, since for any 
$\xi\in\C\backslash\{0\}$, $M$ and $\xi M$ yield the same $W$. One should
therefore identify the space of degree 1 rational maps with the projective
unimodular group $PSL(2,\C)=SL(2,\C)/\Z_2$ \cite{sphere}. 
The most convenient parametrization for our
purposes is
\beq
W(z)=U\odot\left(\frac{\beta}{z-\gamma}\right),
\eeq
with $U\in PSU(2)=SU(2)/\Z_2\cong SO(3)$, $\beta\in (0,\infty)$ and $\gamma
\in\C$.

It remains to substitute $W$ back into the Einstein equation for $\Omega$.
Since changing $U$ merely alters the global internal orientation of $W$, 
$\Omega$ must be independent of $U$, which we may set to ${\Bbb I}$ without
loss of generality. The result, as found by Comtet and Gibbons \cite{comgib}
is
\beq
\Omega(z,\ol{z})=\frac{1}{(\beta^2+|z-\gamma|^2)^\lambda},
\eeq
where $\lambda=8\pi G\mu^2$. If $\lambda<\frac{1}{2}$, the corresponding
spatial metric $h=\Omega^2\, dz\, d\ol{z}$ is asymptotically conical, with
a deficit angle $\delta=4\pi\lambda$, and the singular tip replaced by a
region of smooth distortion centred on $z=\gamma$. If $\lambda=\frac{1}{2}$,
space is asymptotically cylindrical, with one closed, rounded end. If
$\lambda>\frac{1}{2}$, the volume of space is finite (volume$
=\pi\beta^2/(2\lambda-1)$), and except when $\lambda=1$, the boundary of space
$z=\infty$ is replaced by a conical singularity of deficit angle $4\pi(1-
\lambda)$ (or rather ``surfeit angle'' $4\pi(\lambda-1)$ if $\lambda>1$)
lying at finite proper distance\footnote{The authors are grateful to G.W.
Gibbons for pointing this out.}. In the special case $\lambda
=1$, space is uniformly spherical, $z-\gamma$ being a stereographic coodinate.

In summary, by $M_1$ we mean the space of pairs
\beq
\left(W:z\mapsto U\odot\left(\frac{\beta}{z-\gamma}\right),
      h:z\mapsto \frac{dz\, d\ol{z}}{(\beta^{2}+|z-\gamma|^2)^{2\lambda}}
\right)
\eeq
parametrized by $(U,\beta,\gamma)\in PSU(2)\times(0,\infty)\times\C$, all of
which are static solutions of the Einstein-$\CP$ equations.

\section{The restricted action}
\label{sec:action}

It was proved in \cite{comgib} that the static solutions in $M_1$ saturate
a lower bound on deficit angle $\delta$ analogous to the Bogomol'nyi bound
in the non-gravitating system ($\delta$ is an effective, albeit nonlocal, 
measure of the total gravitational energy of the system. In our case it 
coincides with Thorne's C-energy \cite{thorne} since the metric $\eta=dt^2
-h$ has circular symmetry). In the absence of
gravity, it is obvious that time dependence can only increase energy, since
kinetic energy is manifestly non-negative. It is not clear that the same 
statement holds for $\delta$ in our case, although this is conjectured in
\cite{comgib}. The geodesic approximation is usually justified by observing
that energy conservation forces the field to stay close to $M_n$ for low
soliton speeds \cite{man}. Rigorous analysis backs up this physical intuition
\cite{stuvor,stumon}. In applying the same approximation in the presence of 
gravity, we are being somewhat more speculative. Nevertheless, given the 
conjecture above, and previous applications of the approximation to 
gravitating systems \cite{gib}, the collective coordinate approximation seems
sensible, and worthy of investigation.

To obtain the restricted action $I|$, we substitute into $I$ a field and metric
whose time dependence is of the form
\beq
W(t,z)=U(t)\odot\frac{\beta(t)}{z-\gamma(t)},\qquad
\eta(t,z)=dt^2-\frac{dz\, d\ol{z}}{(\beta(t)^2+|z-\gamma(t)|^2)^{2\lambda}}.
\eeq
One finds that
\beq
I|=\int dt\int dz\, d\ol{z}\, \Omega^2
\left[\frac{S_h}{16\pi G}+2\mu^2\frac{|\dot{W}|^2}{(1+|W|^2)^2}
-\frac{2\mu^2}{\Omega^2}\frac{\cd_i W\cd_i\ol{W}}{(1+|W|^2)^2}\right]
=:\int dt\, [L_1+L_2+L_3]
\eeq
where $S_\eta=S_{dt\otimes dt}-S_h=-S_h$ has been used.
Now $L_3$ is (up to a constant factor) the potential energy of the 
non-gravitating system, so this is manifestly constant on $M_1$
(in fact $L_3=-4\pi\mu^2$). An appeal to the local Gauss-Bonnet Theorem
\cite{wil}, or a
 straightforward calculation using the formulae
\bea
re^{i\theta}&=&z-\gamma \nonumber \\
S_h&=& -\frac{1}{r\Omega^2}\frac{d\, }{dr}\left(\frac{r}{\Omega}
\frac{d\Omega}{dr}\right)
\eea
establishes that $L_1$ is also constant ($L_1=\lambda/4G$). 
Hence, both $L_1$ and $L_3$ may
be discarded from $I|$, leaving a reduced action equivalent to a geodesic
problem on $M_1$. Denoting the 6 moduli collectively by $q^a$, so that
$\dot{W}=\dot{q}^a\cd W/\cd q^a$ one finds that
\beq
I|=2\mu^2\int dt\, g_q(\dot{q},\dot{q})
\eeq
where $g$ is the following Riemannian metric on $M_1$:
\beq
\label{*}
g=\left[{\rm Re}\int\, dz\, d\ol{z}\, \frac{|z-\gamma|^4}
{(\beta^2+|z-\gamma|^2)^{2(1+\lambda)}}
\frac{\cd W}{\cd q^a}\frac{\cd \ol{W}}{\cd q^b}\right]\, dq^a\, dq^b.
\eeq

It remains to compute the metric coefficients $g_{ab}$. Recall that $M_1$
is diffeomorphic to $PSU(2)\times(0,\infty)\times\C$. The metric $g$ is 
invariant under changes of $U$ (global internal rotations), so it suffices to
evaluate $g$ in the special case $U={\Bbb I}$. Elsewhere, g will follow
from the action of the isometry group. A convenient coordinate chart on 
$PSU(2)$ may be defined as follows: use the canonical identifcation of $SU(2)$
with the unit 3-sphere $S^3\subset\R^4$, and identify $PSU(2)$ with the
upper hemisphere; project the upper hemisphere (minus boundary)
vertically onto the open equatorial 3-ball, to obtain a 3-vector $\vvec$:
\beq
U=\left(\begin{array}{cc}
\sqrt{1-|\vvec|^2}+iv_3 & v_2+iv_1 \\
v_2-iv_1 & \sqrt{1-|\vvec|^2}-iv_3\end{array}\right),\qquad |\vvec|<1.
\eeq
Note that $\vvec=\zv\Leftrightarrow U={\Bbb I}$. One then finds that
$$
\nonumber
\left.\frac{\cd W}{\cd v_1}\right|_{\vvec=\zv}=
i\left[1-\frac{\beta^2}{(z-\gamma)^2}\right],\qquad
\left.\frac{\cd W}{\cd v_2}\right|_{\vvec=\zv}=
1+\frac{\beta^2}{(z-\gamma)^2},\qquad
\left.\frac{\cd W}{\cd v_3}\right|_{\vvec=\zv}=
\frac{2i\beta}{z-\gamma},
\nonumber
$$
\beq
\left.\frac{\cd W}{\cd \beta}\right|_{\vvec=\zv}=
\frac{1}{z-\gamma},\qquad
\left.\frac{\cd W}{\cd \gamma}\right|_{\vvec=\zv}=
\frac{\beta}{(z-\gamma)^2}.
\eeq
Substituting these expressions into (\ref{*}), it is immediately apparent that
all $g_{ab}$ are independent of $\gamma$, which may thus be set to $0$. The
integrals are easily computed. If $\lambda\leq\frac{1}{2}$ neither $g_{v_1v_1}$
nor $g_{v_2v_2}$ exists, the corresponding integrals being divergent. It
follows that $v_1$ and $v_2$ are frozen, and the metric (at $v_3=0$) is
\beq
\label{wc}
g|_{{\tiny\scriptscriptstyle \vvec=\zv}}=
\frac{\pi}{(2\lambda+1)\beta^{4\lambda}}\left[
d\gamma\, d\ol{\gamma}+\frac{1}{2\lambda}(d\beta^2+4\beta^2dv_3^2)\right].
\eeq
If $\frac{1}{2}<\lambda\leq 1$, all $g_{ab}$ are finite, so all 6 moduli are
free to change with time. In this case,
\beq
\label{sc}
g|_{\tiny\scriptscriptstyle
\vvec=\zv}=\frac{\pi}{(2\lambda+1)\beta^{4\lambda}}\left[
d\gamma\, d\ol{\gamma}+\frac{1}{2\lambda}(d\beta^2+4\beta^2(C(\lambda)dv_1^2
+C(\lambda)dv_2^2+dv_3^2))+2\beta(d\gamma_1\, dv_2-d\gamma_2\, dv_1)\right]
\eeq
where $C(\lambda)=(2\lambda^2-\lambda+1)/(4\lambda-2)$ and $\gamma=\gamma_1
+i\gamma_2$. To globalize these formulae (remove the condition $U={\Bbb I}$)
we may define a left invariant basis $\{\sigma_a:a=1,2,3\}$ on $T^*SU(2)$ by
expanding the left invariant one form $U^{-1}dU$ in a basis for the Lie algebra
$su(2)$. We choose
\beq
U^{-1}dU=\sigvec\cdot\left(\frac{i}{2}\tauvec\right),
\eeq
$\{\tau_a:a=1,2,3\}$ being the Pauli spin matrices. The connexion with the
previous basis is simple:
\beq
\sigma_a|_{\tiny\scriptscriptstyle\vvec=\zv}=
2dv_a|_{\tiny\scriptscriptstyle\vvec=\zv}.
\eeq
The formulae (\ref{wc}), 
(\ref{sc}) are easily re-written using 
$\displaystyle{{\sigma_a}}$, and then hold equally true away from $\vvec=\zv$ by $PSU(2)$ left
translation invariance of $g$. Hence,
\beq
\label{***}
g=\frac{\pi}{(2\lambda+1)\beta^{4\lambda}}\left[
d\gamma\, d\ol{\gamma}+\frac{1}{2\lambda}(d\beta^2+\beta^2\sigma_3^2)\right]
\eeq
if $0<\lambda\leq\frac{1}{2}$, while
\beq
\label{****}
g=\frac{\pi}{(2\lambda+1)\beta^{4\lambda}}\left[
d\gamma\, d\ol{\gamma}+\frac{1}{2\lambda}(d\beta^2+\beta^2(C(\lambda)
(\sigma_1^2+\sigma_2^2)
+\sigma_3^2))+\beta(d\gamma_1\, \sigma_2-d\gamma_2\, \sigma_1)\right]
\eeq
if $\frac{1}{2}<\lambda\leq 1$. In the case of weak coupling, once the boundary
value of $W$ is specified, $U$ must remain in the stabilizer of this boundary
value, a $SO(2)$ subgroup of $PSU(2)\cong SO(3)$. So the lump may spin, but
only about its own axis. Geodesic motion takes place on a single leaf
$M_1^{W(\infty)}$ of a foliation of $M_1$ by 4-manifolds parametrized by the 
boundary value $W(\infty)\in\CP$. 
More general rotational dynamics is possible in the 
regime of strong coupling. Note that $C(1)=1$, so in the case of spherical
spatial geometry, $g$ is exceptionally symmetric, as one would expect.

\section{Geodesics in the case of weak coupling}
\label{sec:geo}

For the sake of simplicity, we henceforth assume that $0<\lambda<\frac{1}{2}$.
One may, without loss of generality, assume that the frozen boundary value is
$W(\infty)=0$. Then $U=\exp(i\psi\tau_3/2)$, and
\beq
W(z)=\frac{\beta e^{i\psi}}{z-\gamma}.
\eeq
This leaf $M_1^0$ of $M_1$ is, like every other,
 diffeomorphic to $\R^4\backslash\R^2$. It has
metric
\beq
\label{full}
g=\frac{\pi}{(2\lambda+1)\beta^{4\lambda}}\left(d\gamma\, d\ol{\gamma}+
\frac{d\beta^2+\beta^2\, d\psi^2}{2\lambda}\right).
\eeq
We remark in passing that $(M_1^0,g)$ is conformally flat.

The two-dimensional submanifold
 on which $\gamma=0$, call it $\Sigma$, is isometric to a cone
of deficit angle $4\pi\lambda$ (with its tip, $\beta=0$, missing). This cone
is a totally geodesic submanifold since it is the fixed point set of the
discrete group of isometries $\{{\rm Id},\gamma\mapsto-\gamma\}$. It 
immediately follows that $(M_1^0,g)$ is geodesically incomplete, a property
found to be universal in the non-gravitating model on compact physical
space \cite{sad}. Generically, geodesics on $\Sigma$ miss the singularity
$\beta=0$ -- they emerge from and return to the asymptotic region 
$\beta\ra\infty$ -- and may be extended forever in the future and past.
In interpreting such a geodesic one must take care to account for the $\beta$
dependence of the metric on physical space, $h$. Given a fixed pair of points
in $\CP$ (not coincident), the proper distance between their preimages under
$W$ scales as $\beta^{1-2\lambda}$, so for $0<\lambda<\frac{1}{2}$ it is
correct to regard $\beta$ as a measure of the lump's width. Hence, the generic
geodesic on $\Sigma$
corresponds to a spinning
lump shrinking to some minimum width, then spreading out
indefinitely, spinning ever more slowly as it expands. There are also 
completely irrotational geodesics which hit $\beta=0$ in finite time, 
corresponding to the lump collapsing to zero width -- 
the geodesic approximation
predicts that self-gravitating lumps are unstable. 

Returning to the full four-dimensional geodesic problem,
the geodesic equation for $\beta$ is
\beq
\label{**}
\ddot{\beta}-\frac{2\lambda}{\beta}\dot{\beta}^2=
-4\lambda^2\beta^{8\lambda-1}\left[|p|^2-\frac{(1-2\lambda)}{\beta^2}J^2
\right],
\eeq
$p=\beta^{-4\lambda}\dot{\gamma}$ and 
$J=\beta^{2(1-2\lambda)}\dot{\psi}/2\lambda$ being the constant momenta
conjugate to the cyclic variables $\ol{\gamma}$ and $\psi$ respectively.
Clearly, the lump maintains constant ``velocity'' $\dot{\gamma}$ if and only
if $\beta$ remains constant, and in this case $\dot{\psi}$ remains constant
too. Hence we may obtain translating-spinning solutions with constant
$\beta$ (shape), $\dot{\gamma}$ (velocity) and $\dot{\psi}$ (angular
velocity) by setting the right hand side of (\ref{**}) to zero. Given an 
initial velocity $\dot{\gamma}(0)=v\in\C$, the lump translates undistorted
with $\beta=\beta_0$, $\dot{\psi}=\omega$, constant, if and only if
$\beta_0$ and $\omega$ satisfy
\beq
(\beta_0\omega)^2=\frac{4\lambda^2}{1-2\lambda}|v|^2.
\eeq
The narrower the lump, the faster it must spin in order to travel undistorted
at a given speed. The generic motion is rather complicated, with speed,
angular velocity and width all oscillating with the same period.

\section{Concluding remarks}

We have derived a collective coordinate approximation for the slow motion of
a single self-gravitating $\CP$ lump by restricting the Einstein-Hilbert-$\CP$
action to the degree 1 moduli space of Comtet and Gibbons, $M_1$. This
approximation takes the form of a geodesic problem either on a single leaf
of a foliation of $M_1$ by 4-manifolds (weak coupling) or on all $M_1$
(strong coupling). The physical interpretation of this is that when 
gravitational coupling is weak, the boundary of space lies at infinite proper
distance, so the boundary value of the field remains frozen. Every leaf of
the foliation is geodesically incomplete, leading us to predict that
self-gravitating lumps are unstable to collapse. 

It would be interesting to investigate the general $n$-lump case along 
similar lines. Here it will be impossible to find closed formulae for $g$
analogous to (\ref{***}) and (\ref{****}). 
However, one should still be able to obtain 
qualitative information (on completeness of $(M_n,g)$, for example) and
identify especially symmetric geodesics. One could also investigate single lump
motion at strong coupling, using the more complicated metric (\ref{****}),
although this regime is presumably far from real cosmological applications
(the lump may be regarded as a slice through an infinitely long, straight
cosmic string).
A rigorous justification for the approximation similar to the work of
Stuart \cite{stuvor,stumon}, seems well beyond reach. Numerical tests of the
approximation would therefore be useful.

%\vspace{1cm}
%\noindent
%{\bf Acknowledgement:} The authors are grateful to G.W. Gibbons for pointing
%out an error in an earlier version of this paper.

\end{document}